\documentclass[%
jmp,
 amsmath,amssymb,
 reprint,%
]{revtex4-1}

\usepackage{graphicx}
\usepackage{dcolumn}
\usepackage{bm}

\usepackage[utf8]{inputenc}
\usepackage[T1]{fontenc}
\usepackage{mathptmx}
\usepackage{etoolbox}

\usepackage[final]{changes}

\usepackage{color}
\definecolor{light}{gray}{0.50}
\definecolor{heavy}{gray}{0.35}
\definecolor{black}{gray}{0.0}
\definecolor{dgreen}{rgb}{0.0,0.5,0}
\definecolor{dred}{rgb}{0.9959,0,0}
\definecolor{green}{rgb}{0.0,0.99599,0.0}
\definecolor{purple}{rgb}{0.6,0.0,0.4}

\definechangesauthor[name=Georg Sebastian, color=red]{GSV}
\definechangesauthor[name=Young-Ha, color=magenta]{YHK}
\definechangesauthor[name=Ulrich, color=blue]{UA}
\definechangesauthor[name=Gergely, color=green]{GB}
\definechangesauthor[name=Günther, color=purple]{GZ}

\makeatletter
\def\@email#1#2{%
 \endgroup
 \patchcmd{\titleblock@produce}
  {\frontmatter@RRAPformat}
  {\frontmatter@RRAPformat{\produce@RRAP{*#1\href{mailto:#2}{#2}}}\frontmatter@RRAPformat}
  {}{}
}%
\makeatother

\newcommand{\eps}{\varepsilon}

\def\ex{\mathbf{e}_{x}}
\def\ey{\mathbf{e}_{y}}
\def\ez{\mathbf{e}_{z}}

\def\pvh{\mathbf{p}_h}
\def\uv{\mathbf{u}}
\def\vv{\mathbf{v}}

\def\xv{\mathbf{x}}
\def\Xv{\mathbf{X}}
\def\kv{\mathbf{k}}

\def\kvhb{\mathbf{k}_{h,\beta}}
\def\khb{k_{h,\beta}}
\def\kvp{\dot{\mathbf{k}}}

\def\cg{\mathbf{c}_g}
\def\cgb{\mathbf{c}_{g,\beta}}
\def\cgh{\mathbf{\hat{c}}_g}

\def\Vv{\mathbf{V}}
\def\Uv{\mathbf{U}}

\def\Tv{\mathbf{T}}
\def\Mv{\mathbf{M}}

\def\vua{\langle \uv \rangle}

\def\pia{\langle \pi \rangle}

\newcommand{\rhobar}{\overline{\rho}}

\def\tbar{\overline{\theta}}

\def\cgh{\mathbf{\hat{c}}_g}

\def\Ac{\mathcal{A}}
\def\Nc{\mathcal{N}}

\def\Gc{\mathcal{G}}
\def\Hc{\mathcal{H}}

\newcommand{\vGc}{{\boldsymbol{\mathcal{G}}}}
\newcommand{\vHc}{{\boldsymbol{\mathcal{H}}}}

\def\oh{\hat{\omega}}

\def\Rhobarnull{{\overline{R}^{(0)}}}

\newcommand{\Pbarnull}{\overline{P}^{(0)}}

\newcommand{\pibarnull}{{\overline{\Pi}^{(0)}}}

\def\tbarnull{{\overline{\Theta}^{(0)}}}
\def\tbaralpha{\overline{\Theta}^{(\alpha)}}

\def\Vv{\mathbf{V}}
\def\Uv{\mathbf{U}}

\newcommand{\Ord}{\mathcal{O}}

\newcommand{\ds}{\ensuremath\displaystyle}

\newcommand{\mySum}[3]{
  \sum\limits_{#1=#2}^{#3}
}

\newcommand{\fld}[3]{{#1}_{#2}^{(#3)}}

\newcommand{\Ufld}[2]{\Uv_{#1}^{(#2)}}

\begin{document}

\preprint{AIP/123-QED}

\title[Gravity Waves]{Multi-Scale Dynamics of \replaced[id=UA]{the Interaction Between Waves and Mean Flows}{Gravity Waves}: \\
       From \added[id=UA]{Nonlinear WKB} Theory to \replaced[id=UA]{Gravity-Wave Parameterizations}{Application}\\ in Weather and Climate Models}
\author{U. Achatz}
\author{Y.-H. Kim}\altaffiliation[now at ]{Research Institute of Basic Sciences, Seoul National University, South Korea}
\author{G.S. Völker}
\affiliation{ 
Institut für Atmosphäre und Umwelt, Goethe Universität Frankfurt, Germany
}

\date{\today}

\begin{abstract}
The interaction between small-scale \replaced[id=UA]{waves}{gravity waves (GW)} and a larger-scale flow can be described by a multi-scale theory that forms the basis for a new class of parameterizations of subgrid-scale \replaced[id=UA]{gravity waves (GW)}{GWs} in weather and climate models. The development of this theory is reviewed here. It applies to all interesting regimes of atmospheric stratification, i.e. also to moderately strong stratification as occurring in the middle atmosphere, and thereby extends classic assumption for the derivation of quasi-geostrophic theory. At strong wave amplitudes a fully nonlinear theory arises that is complemented by a quasilinear theory for weak GW amplitude. The latter allows the extension to a spectral description that forms the basis of numerical implementations that avoid instabilities due to caustics, e.g. from GW reflection. Conservation properties are discussed, for energy and potential vorticity, as well as conditions under which a GW impact on the larger-scale flow is possible. The numerical implementation of the theory for GW parameterizations in atmospheric models is described, and the consequences of the approach are discussed, as compared to classic GW parameterizations. Although more costly than the latter, it exhibits significantly enhanced realism, while being considerably more efficient than an approach where all relevant GWs are to be resolved. \added[id=UA]{The reported theory and its implementation might be of interest also for the efficient and conceptually insightful description of other wave-mean interactions, including those where the formation of caustics presents a special challenge.}
\end{abstract}

\maketitle


\section{\label{sec_intro}Introduction}

With horizontal and vertical wavelengths down to at most 1km and 100m respectively, \added[id=UA]{mesoscale atmospheric waves such as} internal gravity waves (GWs) will not all be simulated explicitly by operational climate models within the foreseeable future. However, without taking their influence into account, climate models miss essential circulation aspects even on the planetary scale \cite{Fritts+Alexander2003,Kimetal2003,Alexanderetal2010}. Hence they must be parameterized, requiring a solid theory for the interaction between \replaced[id=UA]{the waves}{GWs} and a mean flow. Corresponding studies have led in recent years to the emergence of a new class of GW parameterizations (GWP). The present review is to give an overview on these developments, from theoretical investigations to the implementation of a new GWP into a state-of-the-art climate model. For this purpose we first review the dynamics of large-amplitude, locally monochromatic GWs, then discuss spectra of weak-amplitude GWs, next describe the GW impact in general on the mean flow, touch on conservation properties, and finally give a sketch of the hence resulting numerical developments.

\section{\label{sec_largeamp}Large-Amplitude Locally Monochromatic waves}

\cite{Bretherton1966,Grimshaw1975b,Andrews+McIntyre1978a,Andrews+McIntyre1978b} have done fundamental studies of the interaction between GWs and mean flow that have more recently been extended by \cite{Achatz2010,Achatzetal2017,Achatz2022} to consider a wider range of atmospheric stratification, higher GW harmonics, and to give a deepened account on the conditions for GW impacts on the mean flow. Whatever the considered GW amplitudes, we will consider GWs in interaction with a synoptic-scale flow in a hydrostatic reference atmosphere, with profiles $\bar\theta (z), \bar\rho(z), \bar\pi(z)$ of potential temperature, density, and Exner pressure, respectively, that only depend on altitude $z$. We also assume an $f$-plane with a constant Coriolis parameter. One can assume that the ratio between Coriolis frequency $f$ and Brunt-Vaisala frequency $N = \sqrt{(g/\bar\theta) d_z \bar\theta}$, with $g$ the gravitational acceleration, is $f/N = \Ord [\varepsilon^{(5-\alpha)/2}]$, where $\alpha =0,1$ denotes moderately strong or weak stratification, and $\varepsilon = \Ord(1/10)$ is a small parameter. Observations \cite{Nastrom+Gage1985,Calliesetal2014} indicate that, in the spectrum of GWs, most energy is carried by the waves that are in scale just below the synoptic scale, which we assume to be resolved by climate models of interest. A careful analysis then turns out that, with $T_{00}$ a typical atmospheric temperature, representative horizontal and vertical length scales and a representative time scale of such GWs are 
\begin{equation}
    (L_w, H_w) = \left[\varepsilon^{(2+\alpha)/2},\varepsilon^{7/2}\right] \sqrt{R T_{00}} /f \qquad T_w = 1/f
\end{equation}
whence the wind scales are obtained in the large-amplitude case as advective wind scales $(U_w,W_w) = (L_w,H_w)/T_w$. Waves of such amplitude are close to locally inducing static instability, i.e. negative vertical derivatives of total potential temperature, which would cause them to break. The corresponding mean-flow synoptic scales are $(L_s,H_s,T_s) = (L_w,H_w,T_w)/\eps$, i.e. $\varepsilon$ is our scale-separation parameter. Note that the synoptic-scale wind scales are $(U_s,W_s) = (U_w,W_w)$ so that $U_s/(L_s f) =\varepsilon$, i.e. $\varepsilon$ also is the Rossby number of the synoptic-scale flow \cite{Achatzetal2017,Achatz2022}.

The next step is to use the wave wind, length, and time scales, and $T_{00}$, to non-dimensionalize the equations of motion 
\begin{eqnarray}
\label{eq_hor_mom_d}
  D_t \uv + f\ez \times \uv
  &=&
  -c_p \theta \nabla_h \pi\\
  D_t w
  &=&
  - c_p \theta \partial_z \pi - g \\
  D_t \theta
  &=&
  0 \\
  D_t \pi + \frac{R}{c_V} \pi \nabla\cdot\vv
  &=& 0
  \label{eq_exner_pr_d}
\end{eqnarray}
leading to
\begin{eqnarray}
\label{eq_hor_mom}
  \eps^{2+\alpha} \left(D_t \uv + f_0\ez \times \uv\right)
  &=&
  - \frac{c_p}{R} \theta \nabla_h \pi\\
  \eps^7 D_t w
  &=&
  - \frac{c_p}{R} \theta \partial_z \pi - \eps \\
  D_t \theta
  &=&
  0 \\
  D_t \pi + \frac{R}{c_V} \pi \nabla\cdot\vv
  &=& 0
  \label{eq_exner_pr}
\end{eqnarray}
where $D_t = \partial_t + \vv\cdot\nabla$ indicates the material derivative, $\uv$ and $w$ are the horizontal and vertical components of the total wind $\vv$, respectively. $c_p$ and $c_V = c_p - R$ are the specific heat capacities at constant pressure and volume, respectively, with $R$ the ideal gas constant of dry air. $f_0 = 1$ is a non-dimensional placeholder for the Coriolis frequency. We then introduce slow variables $(\Xv,T) = \eps (\xv,t)$, and insert into the non-dimensional equations the WKB expansions for a superposition of a hydrostatic reference atmosphere, a synoptic-scale flow, and a wave field with its higher harmonics,
\begin{eqnarray}
  \label{equation_v_expansion}
  \vv &=&
    \mySum{j}{0}{\infty}
    \eps^j
    \fld{\Vv}{0}{j}
    (\Xv,T)\nonumber\\
  &&+
    \Re
    \mySum{\beta}{1}{\infty}
    \mySum{j}{0}{\infty}
    \eps^j
    \fld{\Vv}{\beta}{j}
    (\Xv,T)
    e^{i\beta\phi(\Xv,T)/\eps}\\
  \label{equation_theta_expansion}
  \theta &=&
  \sum_{j=0}^\alpha \eps^j \overline{\Theta}^{(j)}(Z)
  + \eps^{1+\alpha}
    \mySum{j}{0}{\infty}
    \eps^j
    \fld{\Theta}{0}{j}
    (\Xv,T)
    \nonumber\\
  && + \eps^{1+\alpha}
    \Re
    \mySum{\beta}{1}{\infty}
    \mySum{j}{0}{\infty}
    \eps^j
    \fld{\Theta}{\beta}{j}
    (\Xv,T)
    e^{i \beta\phi(\Xv,T)/\eps}\\
  \label{equation_pi_expansion}
  \pi &=&
    \sum_{j=0}^\alpha \eps^j \overline{\Pi}^{(j)}(Z)
  + \eps^{1+\alpha}
    \mySum{j}{0}{\infty}
    \eps^j
    \fld{\Pi}{0}{j}
    (\Xv,T)
    \nonumber\\
  &&+ \eps^{2+\alpha}
    \Re
    \mySum{\beta}{1}{\infty}
    \mySum{j}{0}{\infty}
    \eps^j
    \fld{\Pi}{\beta}{j}
    (\Xv,T)
    e^{i \beta\phi(\Xv,T)/\eps}
\end{eqnarray}
where $\overline{\Theta}^{(j)}$ and $\overline{\Pi}^{(j)}$ are due to the reference atmosphere, all terms proportional to the phase factors $\exp{i\beta\phi/\eps}$ are contributions from the wave (subscript $\beta = 1$ for the basic wave, and $\beta \ge 2$ for its $\beta$th higher harmonic), and the rest constitutes the synoptic-scale part (subscript 0). Both the wave amplitudes and the synoptic-scale flow are only slowly varying in space and time, as are the local wavenumbers $\beta \kv = \beta \nabla \phi/\eps = \beta \nabla_\Xv \phi = \left(\ex\partial_X + \ey\partial_Y + \ez\partial_Z\right)\phi$ and frequencies $\beta \omega = - \beta \partial_t \phi/\eps = -\beta \partial_T \phi$. The leading orders of the expansions follow from a dimensional analysis of the basic equations. For the synoptic-scale flow they agree in the weak-stratification case $\alpha = 1$ with standard quasigeostrophic scaling \cite[e.g.][]{Pedlosky1987}.

Inserting (\ref{equation_v_expansion}) - (\ref{equation_pi_expansion}) into (\ref{eq_hor_mom}) - (\ref{eq_exner_pr}) and sorting by terms with equal powers in $\eps$ and in the phase factor $\exp{i\phi/\eps}$ one finds from the leading orders in $\eps$ that the mean flow is to leading order horizontal, and that it is in hydrostatic and geostrophic balance, i.e.
\begin{eqnarray}
  \label{equation_mf_hor}
    \fld{W}{0}{0} &=& 0\\
  \label{equation_hydrost}
    \partial_Z \fld{\Pi}{0}{0}
    &=& 
    \frac{R/c_p}{\tbarnull} \left[\fld{B}{0}{0} - \alpha (\tbaralpha/\tbarnull)^2\right]\\
    \label{equation_geost}
    f_0 \ez\times \fld{\Uv}{0}{0} &=& - \frac{c_p}{R} \tbarnull \nabla_{\Xv,h} \fld{\Pi}{0}{0}
\end{eqnarray}
with $\fld{B}{0}{0} = \fld{\Theta}{0}{0}/\tbarnull$ a non-dimensional synoptic-scale buoyancy, and $\nabla_{\Xv,h} = \ex \partial_X + \ey \partial_Y$ the horizontal gradient operator in the slow spatial variables.

The leading-order results for the wave field are that frequency and wave number satisfy a dispersion relation $\omega = \Omega(\kv,\Xv,T)$ for either geostrophic modes or gravity waves, i.e.
\begin{equation}
\label{eq_dispr_nd}
    \Omega(\kv,\Xv,T) 
    = 
    \left\{
  \begin{array}{ll}
    \kv\cdot\fld{\Uv}{0}{0}, & \hbox{GM}\\
    \kv\cdot\fld{\Uv}{0}{0} \pm \sqrt{\frac{\ds {N_0}^2 k_h^2+f_0^2m^2}{\ds \eps^4 k_h^2 + m^2}}, & \hbox{GW}\\
  \end{array}
  \right.
\end{equation}
where the dependence on $\Xv$ and $T$ enters via that of the mean-flow leading-order horizontal wind $\fld{\Uv}{0}{0}$ and the non-dimensional Brunt-Vaisala frequency $N_0^2 = d_Z \tbaralpha / \tbarnull$. The components of the wavenumber vector are defined so that $\kv = k\ex + l\ey +m\ez$, and $k_h = \sqrt{k^2 + l^2}$ is the absolute magnitude of the horizontal wavenumber $\kv_h = k\ex + l\ey$. The dispersion relations entail the eikonal equations
\begin{eqnarray}
  \label{equation_eikonal_k}
    \left(\partial_T + \cg\cdot\nabla_\Xv\right)\omega
    &=&
    \partial_T \Omega\\
  \label{equation_eikonal_om}
    \left(\partial_T + \cg\cdot\nabla_\Xv\right)\kv
    &=&
    -\nabla_\Xv\Omega
\end{eqnarray}
for the development of wave number and frequency, with $\cg = \nabla_{\kv}\Omega = \left(\ex\partial_k + \ey\partial_l + \ez\partial_m\right)\Omega$ the local group velocity.
The wave amplitudes satisfy the polarization relations
\begin{eqnarray}
    \label{equation_pol_uv}
    \fld{\Uv}{\beta}{0}
    &=&
    \frac{\beta^2\kv_h \oh - i f_0 \ez \times \beta\kv_h}{\beta^2\oh^2-f_0^2} \nonumber \\
    && \times \left(1 - \eps^4 \frac{\beta^2 \oh^2}{N_0^2}\right)
    \frac{\fld{B}{\beta}{0}}{i\beta m}\\
    \label{equation_pol_w}
    \fld{W}{\beta}{0}
    &=&
    \frac{i\beta \oh}{N_0^2}
    \fld{B}{\beta}{0}\\
    \label{equation_pol_pi}
    \frac{c_p}{R} \tbarnull \fld{\Pi}{\beta}{0}
    &=&
    \left(1 - \eps^4 \frac{\beta^2 \oh^2}{N_0^2}\right) \frac{\fld{B}{\beta}{0}}{i\beta m}
\end{eqnarray}
with $\oh = \omega - \kv\cdot\fld{\Uv}{0}{0}$ the intrinsic frequency. One also finds that for GWs $\fld{B}{\beta}{0} = 0$ if $\beta > 1$, and only for $j>0$ one can have $\fld{B}{\beta}{j} \neq 0$ if $\beta > 1$. Hence GW higher harmonics do not contribute to the leading order, but potentially only to the next but leading order. This is not the case for the GM higher harmonics, but in the following we ignore this option and only consider wave fields without GM contribution.

A central result from the next-order terms in $\eps$ is the wave-action conservation equation 
\begin{equation}
    \label{equation_wact_cons}
    \partial_T \mathcal{A} + \nabla_\Xv\cdot \left( \cg \mathcal{A} \right) = 0
\end{equation}
for the GW wave-action density $\mathcal{A} = E_{gw}/\oh$, with 
\begin{equation}
  \label{equation_egw}
  E_{gw}=
  \frac{\Rhobarnull}{2}
  \left(
    \frac{|\fld{\Uv}{1}{0}|^2}{2}
    + \eps^4 \frac{|\fld{W}{1}{0}|^2}{2}
    +\frac{1}{{N_0}^2}\frac{|\fld{B}{1}{0}|^2}{2}
  \right)
\end{equation}
the GW energy density, where $\Rhobarnull(Z)$ is the leading-order reference-atmosphere density, decaying strongly from the ground to higher altitudes. The derivation of this result uses the geostrophic and hydrostatic equilibrium (\ref{equation_hydrost}) - (\ref{equation_geost})  of the synoptic-scale flow. 
For the leading-order contributions to the synoptic-scale flow one obtains the prognostic equations
\begin{eqnarray}
    &&\left(1 - \alpha\right) \left(\partial_T + \fld{\Uv}{0}{0} \cdot \nabla_{\Xv,h}\right) \fld{\Pi}{0}{0}\nonumber\\
    &&+ \frac{R}{c_V} \frac{\pibarnull}{\Pbarnull} \nabla_\Xv \cdot \left(\Pbarnull \fld{\Vv}{0}{1}\right)
    = 0
  \label{equation_progmf_pi}\\
   &&\left(\partial_T + \fld{\Uv}{0}{0} \cdot \nabla_{\Xv,h}\right) \fld{\Theta}{0}{0} + \fld{W}{0}{1} \tbarnull N_0^2\nonumber\\
    &&=
    - \frac{1}{\Rhobarnull} \nabla_{\Xv,h} \cdot \Tv
    \label{equation_progmf_theta}\\
    &&
    \left(\partial_T + \fld{\Uv}{0}{0} \cdot \nabla_{\Xv,h}\right) \fld{\Uv}{0}{0} + f_0 \ez\times \fld{\Uv}{0}{1} \nonumber\\
    &&=
    - \frac{c_p}{R} \tbarnull \nabla_{\Xv,h} \fld{\Pi}{0}{1}\nonumber\\
    &&\quad - \frac{c_p}{R} \left[\alpha \tbaralpha + \left(1 - \alpha\right) \fld{\Theta}{0}{0}\right] \nabla_{\Xv,h} \fld{\Pi}{0}{0}\nonumber\\
    &&\quad -\frac{1}{\Rhobarnull} \nabla_\Xv \cdot \Mv + \frac{1-\alpha}{\Pbarnull} f_0 \ez\times \Tv
    \label{equation_progmf_hmom}
\end{eqnarray}
where $\Pbarnull = \Rhobarnull \tbarnull$ is the leading-order mass-weighted potential temperature of the reference atmosphere. While there is no direct GW impact on the synoptic-scale Exner pressure, GW entropy-flux and momentum-flux convergence appear in prognostic equations for mean-flow potential temperature and horizontal momentum. With $\cgh = \nabla_\kv \oh = \cg - \fld{\Uv}{0}{0}$ being the intrinsic group velocity, the entropy fluxes are
\begin{equation}
    \Tv 
    = \frac{\Rhobarnull}{2} \Re \left(\fld{\Uv}{1}{0} {\fld{\Theta}{1}{0}}^*\right)
    = \ez\times\kv_h\, \Ac\, \hat{c}_{gz} \frac{N_0^2 f_0}{\oh^2 - f_0^2} \tbarnull
\end{equation}
and the momentum-flux tensor 
\begin{equation}
    \Mv = \frac{\Rhobarnull}{2} \Re \left(\fld{\Vv}{1}{0} {\fld{\Uv}{1}{0}}^* \right)
\end{equation}
has the elements
\begin{eqnarray}
    M_{11} 
    &=& 
    k \Ac \hat{c}_{gx} \frac{\oh^2}{\oh^2 - f_0^2} 
    + l \Ac \hat{c}_{gy} \frac{f_0^2}{\oh^2 - f_0^2}
    \\
    M_{12} = M_{21} &=& k \Ac \hat{c}_{gy} = l \Ac \hat{c}_{gx} \\
    M_{22} 
    &=& 
    l \Ac \hat{c}_{gy} \frac{\oh^2}{\oh^2 - f_0^2} 
    + k \Ac \hat{c}_{gx} \frac{f_0^2}{\oh^2 - f_0^2}\\
    M_{31} &=& \frac{k \Ac \hat{c}_{gz}}{1 - f_0^2/\oh^2} \\
    M_{32} &=& \frac{l \Ac \hat{c}_{gz}}{1 - f_0^2/\oh^2}
\end{eqnarray}
The last GW term in the horizontal momentum equation (\ref{equation_progmf_hmom}) is the so-called elastic term that appears only at moderately strong stratification, i.e. for $\alpha = 0$. It also vanishes in the absence of rotation.

Note that the results above represent a fully nonlinear theory where the nonlinear advection terms turn out to not contribute to the leading orders of the wave equations because the solenoidality property
\begin{equation}
  \label{equation_solenoidality}
    \kv\cdot\fld{\Vv}{\beta}{0}=0
\end{equation}
of the wave velocity field, following directly from the leading-order wave part of the Exner-pressure equation, eliminates self-advection of the wave field to leading order.


\section{\label{sec_weakamp}A Spectrum of Weak-Amplitude Waves}

The eikonal equations (\ref{equation_eikonal_k}) and (\ref{equation_eikonal_om}) can lead to a breakdown of the local monochromaticity assumed in the derivation of the wave-action equation (\ref{equation_wact_cons}), e.g. if an initial locally monochromatic GW field has a spatial dependence of wave number so that neighboring regions have group velocities leading to caustics where rays cross. This calls for a theory describing the dynamics of the superposition of several wave fields. However, because of the nonlinear advection terms, the superposition of wave fields with strong amplitudes close to breaking, as assumed above, does not even allow the derivation of dispersion and polarization relations anymore. In the locally monochromatic case the advection terms do not contribute to the leading orders of the wave equations, because of the solenoidality property (\ref{equation_solenoidality}). In the case of a superposition of several wave fields, however, mutual advection of the different wave fields only does not contribute to the leading order of the equations if the wave amplitudes are sufficiently weak, i.e. when the wave fields in (\ref{equation_v_expansion}) - (\ref{equation_pi_expansion}) are multiplied by a factor $\eps^n$ with, e.g., $n = 1$, termed \emph{weakly nonlinear} case or $n=2$, the \emph{quasilinear} case. One sets
\begin{eqnarray}
  \label{equation_v_expansion_ql}
  \vv &=&
    \mySum{j}{0}{\infty}
    \eps^j
    \fld{\Vv}{0}{j}
    (\Xv,T)\nonumber\\
  &&+
  \eps^n
    \Re
    \sum_\beta
    \mySum{j}{0}{\infty}
    \eps^j
    \fld{\Vv}{\beta}{j}
    (\Xv,T)
    e^{i\phi_\beta(\Xv,T)/\eps}\\
  \label{equation_theta_expansion_ql}
  \theta &=&
  \sum_{j=0}^\alpha \eps^j \overline{\Theta}^{(j)}(Z)
  + \eps^{1+\alpha}
    \mySum{j}{0}{\infty}
    \eps^j
    \fld{\Theta}{0}{j}
    (\Xv,T)
    \nonumber\\
  && + \eps^{n+1+\alpha}
    \Re
    \sum_\beta
    \mySum{j}{0}{\infty}
    \eps^j
    \fld{\Theta}{\beta}{j}
    (\Xv,T)
    e^{i \phi_\beta(\Xv,T)/\eps}\\
  \label{equation_pi_expansion_ql}
  \pi &=&
    \sum_{j=0}^\alpha \eps^j \overline{\Pi}^{(j)}(Z)
  + \eps^{1+\alpha}
    \mySum{j}{0}{\infty}
    \eps^j
    \fld{\Pi}{0}{j}
    (\Xv,T)
    \nonumber\\
  &&+ \eps^{n+2+\alpha}
    \Re
    \sum_\beta
    \mySum{j}{0}{\infty}
    \eps^j
    \fld{\Pi}{\beta}{j}
    (\Xv,T)
    e^{i \phi_\beta(\Xv,T)/\eps}
\end{eqnarray}
where the index $\beta$ indicates different wave fields. Those could also include higher harmonics of another wave field, but due to the weak wave amplitudes these do not contribute to all relevant orders. Take first a locally monochromatic wave field, where the wave amplitudes are non-zero only for a single $\beta$. Inserting (\ref{equation_v_expansion_ql}) -  (\ref{equation_pi_expansion_ql}) into the equations of motion (\ref{eq_hor_mom}) - (\ref{eq_exner_pr}), sorting by terms with equal powers in $\eps$, and separating wave part and mean-flow part by averaging over scales sufficiently longer than the wave scales one retrieves the mean-flow results (\ref{equation_mf_hor}) - (\ref{equation_geost}). From the leading-order wave contributions one again obtains the dispersion relations $\omega_\beta = \Omega(\kv_\beta,\Xv,T)$ for either geostrophic modes or gravity waves, entailing also the eikonal equations, i.e.
\begin{eqnarray}
  \label{equation_eikonal_kb}
    \left(\partial_T + \cgb\cdot\nabla_\Xv\right)\omega_\beta
    &=&
    \partial_T \Omega\\
  \label{equation_eikonal_omb}
    \left(\partial_T + \cgb\cdot\nabla_\Xv\right)\kv_\beta
    &=&
    -\nabla_\Xv\Omega
\end{eqnarray}
With the replacements $\beta(\kv,\oh) \rightarrow (\kv_\beta,\oh_\beta)$ one also obtains the polarization relations (\ref{equation_pol_uv}) - (\ref{equation_pol_pi}).

In the \emph{quasilinear case} $n=2$ one obtains from the next-order terms in $\eps$ the wave-action conservation equation 
\begin{equation}
    \label{equation_wact_cons_ql}
    \partial_T \mathcal{A_\beta} + \nabla_\Xv\cdot \left( \cgb \mathcal{A_\beta} \right) = 0
\end{equation}
for the GW wave-action density $\mathcal{A_\beta} = E_{gw,\beta}/\oh_\beta$, with 
\begin{equation}
  \label{equation_egw_ql}
  E_{gw,\beta}=
  \frac{\Rhobarnull}{2}
  \left(
    \frac{|\fld{\Uv}{\beta}{0}|^2}{2}
    + \eps^4 \frac{|\fld{W}{\beta}{0}|^2}{2}
    +\frac{1}{{N_0}^2}\frac{|\fld{B}{\beta}{0}|^2}{2}
  \right)
\end{equation}
the energy density of the GW field indicated by $\beta$. One then observes that leading and next-order wave terms giving the results above are strictly linear in the wave fields. Hence the superposition (\ref{equation_v_expansion_ql}) - (\ref{equation_pi_expansion_ql}) is a solution as well, with each component satisfying its own set of eikonal equations (\ref{equation_eikonal_omb}) and (\ref{equation_eikonal_kb}) and wave-action equation (\ref{equation_wact_cons_ql}). Defining for this superposition the spectral wave-action density
\begin{equation}
  \label{equation_wact_sp}
    \Nc \left(\Xv,\kv,T\right)= \sum_\beta \Ac_\beta \left(\Xv, T\right) \delta \left[\kv - \kv_\beta \left(\Xv, T\right)\right]
\end{equation}
one can show from the eikonal equations and the wave-action equations that it satisfies the conservation equation
\begin{equation}
  \label{equation_wact_cons_sp}
  \partial_T\Nc + \nabla_\Xv \cdot \left(\cg \Nc\right) + \nabla_\kv \cdot \left(\kvp \Nc\right) = 0
\end{equation}
where we have introduced the more compact notation $\kvp = - \nabla_\Xv \Omega$. This equation indicates that the integral of phase-space wave-action density over the total phase-space volume is conserved. Moreover, $\cg = \nabla_\kv\Omega$ and the definition of $\kvp = -\nabla_\Xv \Omega$ imply that the six-dimensional phase-space velocity is non-divergent,
\begin{equation}
  \label{equation_vpv_ndiv}
    \nabla_\Xv \cdot \cg + \nabla_\kv \cdot \kvp = 0
\end{equation}
Hence the conservation equation (\ref{equation_wact_cons_sp}) can also be written
\begin{equation}
  \label{equation_wact_cons_sp_lag}
  \partial_T \Nc + \cg \cdot \nabla_\Xv \Nc + \kvp \cdot \nabla_\kv \Nc = 0
\end{equation}
In contrast to wave-action density $\Ac = \sum_\beta \Ac_\beta = \int d^3k \Nc$ in position space, the spectral wave-action density is conserved along rays in phase space satisfying $d_T (\xv,\kv) =  (\cg,\kvp)$. We also note that the superposition (\ref{equation_wact_sp}) allows for an arbitrary number of components with arbitrary wavenumbers each, so that effectively spectral wave-action density can also be a truly continuous function of wavenumber.

In the \emph{weakly nonlinear case} $n=1$ the wave-action equation is to be supplemented by the effect of wave-wave interactions, of either GWs with GWs or GWs with GMs. To the best of our knowledge, the corresponding scattering integrals have only been worked out fully for Boussinesq dynamics without mean flow \cite{Hasselmann1966,Edenetal2019}, and first steps for Boussinesq dynamics with non-vanishing mean flows have been taken by \cite{Voelkeretal2021}. In the atmospheric context they have so far been simply ignored.

Both in the quasilinear and in the weakly nonlinear case the prognostic equations for the mean flow are still (\ref{equation_progmf_pi}) - (\ref{equation_progmf_hmom}), but now the contributing fluxes are due to the full spectrum, i.e.
\begin{equation}
\label{eq_entrflux}
    \Tv 
    = \eps^{2n} \int d^3k\,
    \ez\times\kv_h\, \Nc\, \hat{c}_{gz} \frac{N_0^2 f_0}{\oh^2 - f_0^2} \tbarnull
\end{equation}
and the momentum-flux tensor has the elements
\begin{eqnarray}
    M_{11} 
    &=& 
    \eps^{2n} \int d^3k \left(\frac{k \Nc \hat{c}_{gx}}{1 - f_0^2/\oh^2} 
    + \frac{l \Nc \hat{c}_{gy}}{\oh^2/f_0^2 - 1}\right) \\
    M_{12} = M_{21} &=& \eps^{2n} \int d^3k\, k \Nc \hat{c}_{gy} = \eps^{2n} \int d^3k\, l \Nc \hat{c}_{gx} \\
    M_{22} 
    &=& 
    \eps^{2n} \int d^3k \left(\frac{l \Nc \hat{c}_{gy}}{1 - f_0^2/\oh^2} 
    + \frac{k \Nc \hat{c}_{gx}}{\oh^2/f_0^2 - 1}\right) \\
    M_{31} &=& \eps^{2n} \int d^3k \frac{k \Nc \hat{c}_{gz}}{1 - f_0^2/\oh^2} \\
    M_{32} &=& \eps^{2n} \int d^3k \frac{l \Nc \hat{c}_{gz}}{1 - f_0^2/\oh^2}
    \label{eq_momflux_32}
\end{eqnarray}
Note that these fluxes only contribute to leading order in the large-amplitude case. It is known, however, that their effects accumulate over longer times so that they cannot be ignored. The cleanest way to handle this would be to introduce a correspondingly slow time scale. For simplicity we avoid this step and just keep the wave impacts on the mean flow as outlined above. In the following we will use (\ref{eq_entrflux}) - (\ref{eq_momflux_32}) also for the monochromatic large-amplitude case, where it is understood that $n=0$ and that $\Nc (\Xv,\kv,T) = \Ac (\Xv,T) \delta\left[\kv - \kv(\Xv,T)\right]$, with $\kv$ as an argument of $\Nc$ not depending on $\Xv$ and $T$ anymore.

\section{\label{sec_qg}Wave Impact on the Balanced Mean Flow}

Because of the geostrophic and hydrostatic equilibrium (\ref{equation_hydrost}) - (\ref{equation_geost}) the synoptic-scale mean flow satisfies an extended quasigeostrophic theory.  Using these equilibrium conditions, one can derive from the prognostic equations (\ref{equation_progmf_pi}) - (\ref{equation_progmf_hmom}) a single prognostic equation 
\begin{eqnarray}
    &&\left(\partial_T + \Ufld{0}{0} \cdot \nabla_{\Xv,h}\right)
    \fld{P}{0}{0}\nonumber\\
    &&=
    - \partial_X \left(\frac{1}{\Rhobarnull} \nabla_\Xv \cdot \vHc\right)
    + \partial_Y \left(\frac{1}{\Rhobarnull} \nabla_\Xv \cdot \vGc\right)
  \label{equation_prog_qgpv}
\end{eqnarray}
for the quasigeostrophic potential vorticity (QGPV)
\begin{eqnarray}
  \label{equation_qgpv}
    \fld{P}{0}{0}
    &=&
    {\nabla_{\Xv,h}}^2 \left(\frac{c_p}{R} \frac{\tbarnull\fld{\Pi}{0}{0}}{f_0}\right)\nonumber\\
    &&+
    \frac{
      f_0
    }{
      \Rhobarnull
    }
    \partial_Z
    \left[
      \frac{
	\Rhobarnull
      }{
	{N_0}^2
      }
      \partial_Z
      \left(
	\frac{c_p}{R} \tbarnull
	\fld{\Pi}{0}{0}
      \right)
    \right]
\end{eqnarray}
with $\vGc = \int d^3 k\, \cgh k \Nc$ and $\vHc = \int d^3 k\, \cgh l \Nc$ the fluxes of the zonal and meridional components of pseudomomentum $\pvh = \int d^3 k\, \kv_h \Nc$. Inverting QGPV yields the leading-order synoptic-scale Exner-pressure fluctuations $\fld{\Pi}{0}{0}$ whence one can obtain, using geostrophic and hydrostatic equilibrium, the leading-order horizontal wind and potential-temperature fluctuations of the synoptic-scale flow. QGPV is forced by the vertical curl of the pseudomomentum-flux convergences, and in the absence of waves it is conserved.

\section{\label{sec_dim}Summary of the results in dimensional form}

The practitioner needs the results above in their dimensional form. These are as follows: A re-dimensionalization of the GW dispersion relation in (\ref{eq_dispr_nd}), by the substitutions
\begin{eqnarray}
  \oh                           &\rightarrow& \oh T_w = \oh/f \\
  \overline{\Theta}^{(\alpha)}  &\rightarrow&   \left\{
                                                \begin{array}{ll}
                                                \tbar/T_{00}                                & \,\mathrm{if}\, \alpha = 0 \\
                                                \left(\tbar/T_{00} - \tbarnull\right)/\eps  & \,\mathrm{if}\, \alpha = 1
                                                \end{array}
                                                \right. \\
  Z                             &\rightarrow& \eps z/H_w \\
  (k,l,m)                       &\rightarrow& \left[L_w (k,l),H_w m\right] \\
  f_0                           &\rightarrow& f/f\\
  \fld{\Uv}{0}{0}               &\rightarrow& \langle\uv\rangle/U_w
\end{eqnarray}
leads to the dimensional GW dispersion relation
\begin{equation}
  \label{equation_dispr}
  \oh^2 = (\omega - \kv_h\cdot\langle\uv\rangle)^2
  =
  \frac{N^2 k_h^2 + f^2 m^2}{k_h^2 + m^2}
\end{equation}
with $N^2 = (g/\tbar) d_z\tbar$, and the trivial reformulation 
\begin{equation}
  \label{equation_dispr_beta}
  \oh_\beta^2 = (\omega_\beta - \kvhb\cdot\langle\uv\rangle)^2
  =
  \frac{N^2 \khb^2 + f^2 m_\beta^2}{\khb^2 + m_\beta^2}
\end{equation}
for the weakly nonlinear and quasilinear case with a superposition of spectral components. Substituting 
\begin{equation}
    \left(\fld{\Uv}{\beta}{0},\fld{W}{\beta}{0},\fld{B}{\beta}{0},\fld{\Pi}{\beta}{0}\right)
    \rightarrow
    \left(\frac{\uv'_\beta}{U_w},\frac{w'_\beta}{W_w},\frac{\theta'_\beta}{\eps^{1+\alpha} \tbar},\frac{\pi'_\beta}{\eps^{2+\alpha}}\right)
\end{equation}
one obtains from (\ref{equation_pol_uv}) - (\ref{equation_pol_pi}) the dimensional polarization relations
\begin{eqnarray}
    \uv'_\beta
    &=&
    \frac{\kvhb \oh_\beta - i f \ez \times \kvhb}{\oh_\beta^2-f^2} 
    \left(1 - \frac{\oh_\beta^2}{N^2}\right)
    \frac{b'_\beta}{i m_\beta}
    \label{equation_pol_u_dim}\\
    \label{equation_pol_w_dim}
    w'_\beta
    &=&
    \frac{i\oh_\beta}{N^2} b'_\beta\\
    c_p \tbar \pi'_\beta
    &=&
    \left(1 - \frac{\oh_\beta^2}{N^2}\right)  \frac{b'_\beta}{i m_\beta}
    \label{equation_pol_pi_dim}
\end{eqnarray}
where $b'_\beta = g\,\theta'_\beta/\tbar$ is the dimensional buoyancy of the $\beta$th GW component. The corresponding results for the locally monochromatic large-amplitude case are obtained from (\ref{equation_pol_u_dim}) - (\ref{equation_pol_pi_dim}) by dropping the $\beta$-index.

Likewise, we obtain for the locally monochromatic case and for the quasilinear spectral case, respectively, the dimensional GW wave-action equations
\begin{equation}
\label{eq_waveact}
    \partial_t \Ac + \nabla\cdot (\mathbf{c}_g \Ac) = 0 \qquad 
    \partial_t \Ac_\beta + \nabla\cdot (\cgb \Ac_\beta) = 0
\end{equation}
where $(\mathbf{c}_g,\cgb) = (\nabla_\kv \omega,\nabla_\kv \omega_\beta)$ is the GW group velocity, and $(\Ac,\Ac_\beta) = (E_w/\oh,E_{w,\beta}/\oh_\beta)$ the GW wave action, with
\begin{equation}\label{eq_ew_dim}
    (E_w,E_{w,\beta})
    =
    \frac{\rhobar}{2}
    \left(
    \frac{\left|\vv'\right|^2}{2} + \frac{\left|b'\right|^2}{2 N^2},
    \frac{\left|\vv'_\beta\right|^2}{2} + \frac{\left|b'_\beta\right|^2}{2 N^2}
    \right)
\end{equation}
the wave energy. $\rhobar$ is the reference-atmosphere density. 

The wave-action equations together with the dimensional forms
\begin{eqnarray}
  \label{equation_eikonal_k_dim}
    \left(\partial_t + \cg\cdot\nabla_\xv\right)\kv
    &=&
    -\nabla_\xv \Omega\\
  \label{equation_eikonal_kb_dim}
    \left(\partial_t + \cgb\cdot\nabla_\xv\right)\kv_\beta
    &=&
    -\nabla_\xv \Omega
\end{eqnarray}
of the eikonal equations (\ref{equation_eikonal_k}) and (\ref{equation_eikonal_kb}), where
\begin{eqnarray}
\label{eq_dispr_nd}
    \Omega(\kv,\xv,t) 
    &=& 
    \kv\cdot\langle\uv\rangle (\xv,t) \pm \sqrt{\frac{\ds {N}^2 (z) k_h^2+f^2m^2}{\ds k_h^2 + m^2}}
\end{eqnarray}
leads to the spectral wave-action equations
\begin{eqnarray}
  \label{equation_wact_cons_sp_dim}
  \partial_t\Nc + \nabla_\xv \cdot \left(\cg \Nc\right) + \nabla_\kv \cdot \left(\kvp \Nc\right) &=& 0\\
  \label{equation_wact_cons_sp_lag_dim}
  \partial_t \Nc + \cg \cdot \nabla_\xv \Nc + \kvp \cdot \nabla_\kv \Nc &=& 0
\end{eqnarray}
for the spectral wave-action density
\begin{equation}
  \label{equation_wact_sp_dim}
    \Nc \left(\xv,\kv,t\right)= \sum_\beta \Ac_\beta \left(\xv, t\right) \delta \left[\kv - \kv_\beta \left(\xv, t\right)\right]
\end{equation}
in the quasilinear case and $\Nc (\xv,\kv,t) = \Ac (\xv,t) \delta\left[\kv - \kv(\xv,t)\right]$ in the locally monochromatic large-amplitude case. Here as well we have defined $\kvp = - \nabla_\xv\Omega$. In deriving (\ref{equation_wact_cons_sp_lag_dim}) from (\ref{equation_wact_cons_sp_dim}) one again exploits the non-divergence 
\begin{equation}
\label{eq_phspvel_nondiv}
    \nabla_\xv \cdot \cg + \nabla_\kv \cdot \kvp = 0
\end{equation} 
of the phase-space velocity.

The GW impact on the synoptic-scale mean flow, indicated by angle brackets $\langle\dots\rangle$, is captured within synoptic scaling by supplementing the entropy equation by GW entropy-flux convergence,
\begin{equation}
  \label{eq_synpteq_dim}
  \left(\partial_t + \langle\vv\rangle \cdot \nabla\right) \langle\theta\rangle = - \nabla \cdot \langle \uv' \theta'\rangle
\end{equation}
and the horizontal-momentum equation by GW momentum-flux convergence and the elastic term
\begin{eqnarray}
  && \left(\partial_t + \langle\vv\rangle \cdot \nabla\right) \vua + f \ez \times \vua \nonumber\\
  &&= 
  - c_p \langle\theta\rangle \nabla_h \pia - \frac{1}{\rhobar} \nabla \cdot \left(\rhobar \langle\vv' \uv'\rangle\right)
  + \frac{f}{\tbar} \ez \times \nabla \cdot \langle \uv' \theta'\rangle
  \label{eq_hormmomeq_dim}
\end{eqnarray}
where the entropy flux can be obtained from the spectral wave-action density via
\begin{equation}
\label{eq_entrflux_dim}
    \langle \uv' \theta'\rangle 
    = \int d^3k\,
    \ez\times\kv_h\, \Nc\, \hat{c}_{gz} \frac{N^2 f}{\oh^2 - f^2} \frac{\tbar}{\rhobar}
\end{equation}
and the mass-specific momentum-flux tensor has the elements
\begin{eqnarray}
\label{eq_momflux_uu_dim}
    \langle u' u'\rangle 
    &=& 
    \int d^3k \left(\frac{k \Nc \hat{c}_{gx}}{1 - f^2/\oh^2} 
    + \frac{l \Nc \hat{c}_{gy}}{\oh^2/f^2 - 1}\right) \\
    \langle u' v'\rangle = \langle v' u'\rangle &=& \int d^3k\, k \Nc \hat{c}_{gy} = \int d^3k\, l \Nc \hat{c}_{gx} \\
    \langle v' v'\rangle 
    &=& 
    \int d^3k \left(\frac{l \Nc \hat{c}_{gy}}{1 - f^2/\oh^2} 
    + \frac{k \Nc \hat{c}_{gx}}{\oh^2/f^2 - 1}\right) \\
    \langle w' u'\rangle &=& \int d^3k \frac{k \Nc \hat{c}_{gz}}{1 - f^2/\oh^2} \\
    \langle w' v'\rangle &=& \int d^3k \frac{l \Nc \hat{c}_{gz}}{1 - f^2/\oh^2}
    \label{eq_momflux_wv_dim}
\end{eqnarray}
In equations (\ref{eq_synpteq_dim}) - (\ref{eq_hormmomeq_dim}) the mean flow is understood to be its full expansion 
\begin{eqnarray}
  \label{equation_vmean_expansion}
  \langle\vv\rangle &=&
    U_w \mySum{j}{0}{\infty}
    \eps^j
    \fld{\Vv}{0}{j}
    (\Xv,T)
    \\
  \label{equation_thetamean_expansion}
  \langle\theta\rangle &=&
  T_{00} 
  \left[\sum_{j=0}^\alpha \eps^j \overline{\Theta}^{(j)}(Z)
  + \eps^{1+\alpha}
    \mySum{j}{0}{\infty}
    \eps^j
    \fld{\Theta}{0}{j}
    (\Xv,T)
    \right]
    \\
  \label{equation_pimean_expansion}
  \langle\pi\rangle &=&
    \sum_{j=0}^\alpha \eps^j \overline{\Pi}^{(j)}(Z)
  + \eps^{1+\alpha}
    \mySum{j}{0}{\infty}
    \eps^j
    \fld{\Pi}{0}{j}
    (\Xv,T)
\end{eqnarray}
so that (\ref{eq_synpteq_dim}) is in the two leading orders in $\eps$ consistent with (\ref{equation_mf_hor}) and (\ref{equation_progmf_theta}), and that (\ref{eq_hormmomeq_dim}) is in the two leading orders consistent with (\ref{equation_geost}) and (\ref{equation_progmf_hmom}).

The fluxes (\ref{eq_entrflux_dim}) - (\ref{eq_momflux_wv_dim}) are to be coded into weather-forecast and climate models, but the leading-order synoptic-scale mean-flow dynamics can also be expressed by the QG potential-vorticity equation
\begin{equation}
    \left(\partial_t + \vua \cdot \nabla_h\right) P
    =
    - \partial_x \left(\frac{1}{\rhobar} \nabla \cdot \vHc\right)
    + \partial_y \left(\frac{1}{\rhobar} \nabla \cdot \vGc\right)
  \label{eq_pv_igwimp_dim}
\end{equation}
with $\vGc = \int d^3 k\, \cgh k \Nc$ and $\vHc = \int d^3 k\, \cgh l \Nc$ the fluxes of the zonal and meridional components of GW pseudomomentum $\pvh = \int d^3 k\, \kv_h \Nc$, 
\begin{equation}
  \label{eq_pv_dim}
    P = \nabla_h^2 \psi + \frac{1}{\rhobar} \frac{\partial }{\partial z} \left(\rhobar \frac{f^2}{N^2} \frac{\partial \psi}{\partial z} \right)
\end{equation}
and $\psi = c_p \tbar_0 \langle\delta\pi\rangle/f$ the streamfunction, where $\langle\delta\pi\rangle = \eps^{1+\alpha} \fld{\Pi}{0}{0}$ is the leading-order synoptic-scale Exner-pressure fluctuations, and $\tbar_0 = T_{00} \tbarnull$ is the leading-order reference-atmosphere potential temperature. The latter depends on $z$ only in the case with moderately strong stratification ($\alpha = 0$), while it is a constant in the weakly stratified case ($\alpha = 1$). The streamfunction also yields the leading-order synoptic-scale horizontal wind via geostrophic equilibrium,
\begin{equation}
\label{eq_geostr_wind}
    \vua = \ez \times \nabla_h\psi
\end{equation}
and the leading-order synoptic-scale potential temperature fluctuations $\langle\delta\theta\rangle = \eps^{1+\alpha} T_{00} \fld{\Theta}{0}{0}$, via hydrostatic equilibrium,
\begin{equation}
    g \frac{\langle\delta\theta\rangle}{\tbar_0}
    =
    f \frac{\partial \psi}{\partial z}
    +
    \left\{
    \begin{array}{cc}
      \ds g \left(\frac{\tbar - \tbar_0}{\tbar_0}\right)^2 & \mathrm{if}\,\alpha = 1 \\
      \ds - N^2 f \psi / g & \mathrm{if}\,\alpha = 0
    \end{array}
    \right.
\end{equation}

\section{\label{sec_cons}Conservation Properties}

Neither GW energy is conserved nor is QGPV. The interaction between GWs and synoptic-scale mean flow leads to an exchange between the two so that the corresponding conserved quantity comprises contributions from both components.

\subsection{\label{ss_energy}Energy}

We begin with energy. As can be shown in the derivation of the wave-action equation, GW energy
\begin{equation}\label{eq_gwen}
  E_w = \int d^3 k \, \oh \Nc
\end{equation}
satisfies
\begin{equation}\label{eq_energyth_ref_spectral}
  \partial_t E_w = - \nabla \cdot \mathbf{F}_w - \left(\ex \vGc + \ey \vHc\right) \cdot\cdot \nabla \vua
\end{equation}
with
\begin{equation}\label{eq_enflux}
  \mathbf{F}_w = \int d^3 k \, \cg \oh \Nc
\end{equation}
the wave-energy flux. The last term describes the exchange with the synoptic-scale mean flow. The latter has an energy density
\begin{equation}\label{eq_encons_es}
  E_s = \frac{\rhobar}{2} \left[\left|\nabla_h \psi\right|^2 + \frac{f^2}{N^2} \left(\frac{\partial \psi}{\partial z}\right)^2\right]
\end{equation}
where the first part is the kinetic energy density and the second part is the density of available potential energy. As can be derived from the QGPV equation (\ref{eq_pv_igwimp_dim}), it obeys
\begin{align}\label{eq_encons_desdt}
  & \partial_t E_s
  - \nabla_h \cdot \left[\rhobar \psi \partial_t \left( \nabla_h \psi\right)\right] 
  - \partial_z \left(\rhobar \psi \frac{f^2}{N^2} \partial_z\partial_t \psi\right)
  \nonumber\\
  &- \nabla_h \cdot \left(\rhobar \psi \vua P \right)
  =
  \partial_x \left(\psi \nabla \cdot \vHc\right) 
  - \partial_y \left(\psi \nabla \cdot \vGc\right)
  \nonumber\\
  &- \nabla \cdot \left[\vua \cdot \left(\ex \vGc + \ey \vHc\right)\right]
  + \left(\ex \vGc + \ey \vHc\right) \cdot\cdot \nabla \vua
\end{align}
Hence the prognostic equation for the total energy $E_s + E_w$,
\begin{align}
  &\partial_t \left(E_s + E_w\right)
  - \nabla_h \cdot \left[\rhobar \psi \left(\partial_t\partial \nabla_h \psi + \vua P\right)\right]
  \nonumber\\
  &- \partial_z \left(\rhobar \psi \frac{f^2}{N^2} \partial_z\partial_t \psi\right)
  \nonumber\\
  &=
  - \nabla \cdot \bigg[\mathbf{F}_w - \ex \psi \nabla \cdot \vHc + \ey \psi \nabla \cdot \vGc \nonumber\\
  &\qquad\qquad + \vua \cdot \left(\ex \vGc + \ey \vHc\right)\bigg]  
\end{align}
contains only flux terms, so that under suitable boundary conditions the volume-integrated total energy is conserved.

\subsection{\label{ss_pv}Potential Vorticity}

For the derivation of a potential-vorticity conservation property one needs a prognostic equation for pseudomomentum $\pvh = \int d^3k\, \kv_h \Nc$. As can be derived with the help of the wave-action-density equation (\ref{equation_wact_cons_sp_dim}), this prognostic equation is
\begin{equation}\label{eq_pm_eq}
  \left(\partial_t + \vua \cdot \nabla\right) \pvh
  =
  - \nabla \cdot \left(\vGc \ex + \vHc \ey\right) - \nabla_h \vua \cdot \pvh
\end{equation}
The vertical curl of this yields
\begin{equation}
   \left(\partial_t + \vua \cdot \nabla\right) \left(\ez \cdot \nabla \times \pvh \right) 
   =
   - \partial_x \nabla \cdot \vHc + \partial_y \nabla \cdot \vGc
\end{equation}
Comparing with the QGPV equation (\ref{eq_pv_igwimp_dim}) one obtains the conservation equation
\begin{equation}\label{ea_pv_cons_wgw}
  \left(\partial_t + \vua \cdot \nabla\right) \Pi = 0
  \qquad
  \Pi = P - \ez \cdot \nabla \times \frac{\pvh}{\rhobar}
\end{equation}
for an extension $\Pi$ of quasigeostrophic potential vorticity that contains contributions from the synoptic-scale flow, that are linear in the synoptic-scale streamfunction, and the negative of the gravity-wave pseudovorticity $\ez \cdot \nabla \times \pvh/\rhobar$, that is nonlinear in the gravity-wave amplitudes. This conservation property can only be broken by non-conservative effects, e.g. GW sources or dissipative GW breaking.

\subsection{\label{ss_na}Non-Acceleration}

Of considerable consequence for the properties of GW parameterizations is the direct consequence
\begin{equation}\label{ea_pv_nonacc}
  \left(\partial_t + \vua \cdot \nabla\right) P 
  = 
  - \left(\partial_t + \vua \cdot \nabla\right) \left(\ez \cdot \nabla \times \frac{\pvh}{\rhobar}\right)
\end{equation}
of (\ref{ea_pv_cons_wgw}). Hence the leading-order synoptic-scale mean flow is not influenced by GWs if GW pseudovorticity is a Lagrangian invariant of the flow. The latter is the case, e.g., if
\begin{itemize}
    \item GW amplitudes and wavenumbers are steady,
    \item GW pseudovorticity does not vary horizontally, and
    \item the GWs are not affected by sources or sinks.
\end{itemize}
Classic GW parameterizations assume steady-state GW fields and they do not take horizontal variations of the GWs fields into account. Hence they rely exclusively on GW sources and GW breaking as processes leading to a GW impact on the resolved flow.

\section{\label{sec_gwp}Consequences for GW parameterizations}

\subsection{\label{ss_numerics}Numerical Implementation}

\begin{figure}[!t]
\includegraphics[width=0.45\textwidth]{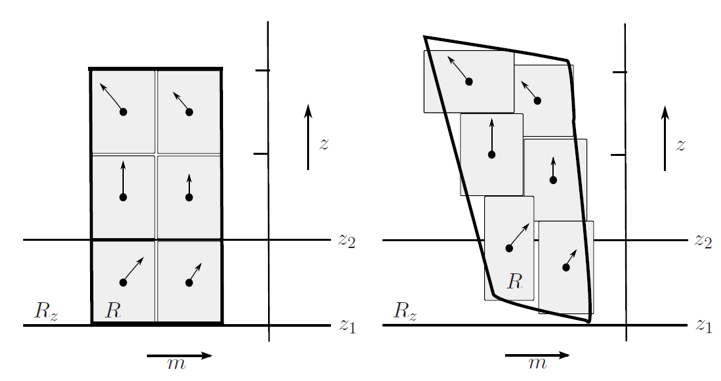}
\caption{\label{fig:rayvol} Illustration of the Lagrangian discretization of the propagation of a phase-space volume, here in the sub-space spanned only by $m$ and $z$. It is subdivided into small rectangular ray volumes. Each ray volume propagates with a mean velocity averaged from the phase-space velocities at the center of the upper and lower edge of the ray volume. Different group velocities $c_{gz}$ at the edges lead to a stretching or squeezing of the ray-volume extent $\Delta z$ in $z$-direction. The extent $\Delta m$ in $m$ direction is then adjusted so that the area $\Delta z \Delta m$ is conserved. The procedure in the $x-k$ and $y-l$ sub-spaces is the same. \replaced[id = UA]{Reproduced from J. Muraschko, M. Fruman, U. Achatz, S. Hickel, and Y. Toledo, Quart. J. R. Met. Soc. 141, 676 (2015), with permission}{Reprinted with permission from \cite{Muraschkoetal2015}}}
\end{figure}
A numerical implementation of spectral wave-action dynamics could either be done in a Eulerian finite-volume formulation, based on (\ref{equation_wact_cons_sp_dim}), or in a Lagrangian approach starting from (\ref{equation_wact_cons_sp_lag_dim}). In the latter \cite{Muraschkoetal2015,Boeloenietal2016,Kimetal2021,Voelkeretal2021} one sub-divides the part of phase space with non-zero wave-action density into rectangular so-called ray volumes. Wave-action density is conserved along trajectories (called rays) satisfying
\begin{equation}
    d_t(\xv,\kv) = (\cg,\kvp)
\end{equation}
Hence, following each point of a ray volume along the ray passing through it one obtains its propagation as a time-dependent volume within which wave-action density $\Nc$ is conserved. Because of the non-divergence (\ref{eq_phspvel_nondiv}) of the phase-space (ray) velocity the volume content of this volume is conserved as well. In the numerical discretization one assumes that each ray volume keeps a rectangular shape, but it is allowed to be stretched and squeezed in a volume-preserving manner. Because (\ref{eq_phspvel_nondiv}) holds separately in the two-dimensional spaces spanned by $x$ and $k$, $y$ and $l$, or $z$ and $m$, i.e.
\begin{eqnarray}
    \partial_x c_{gx} + \partial_k \dot{k} &=& 0\\
    \partial_y c_{gy} + \partial_l \dot{l} &=& 0\\
    \partial_z c_{gz} + \partial_m \dot{m} &=& 0
\end{eqnarray}
this is done so that the areas $\Delta x \Delta k$, $\Delta y \Delta l$ and $\Delta z \Delta m$ are conserved, with $\Delta x$, $\Delta y$, $\Delta z$, $\Delta k$, $\Delta l$, and $\Delta m$ the ray volume extent in the six respective phase-space directions. Fig. \ref{fig:rayvol} illustrates this for the space spanned by $z$ and $m$. Volume deformations away from an original rectangular shape are to be represented by splitting a corresponding volume into a sufficiently large number of rectangular ray volumes.

GW dissipation is simulated by a classic saturation approach based on \cite{Lindzen1981} and adapted by \cite{Boeloenietal2016,Boeloenietal2021} to spectral wave-action dynamics. A static-instability breaking threshold, with a tuning factor close to 1, is determined by the criterion that, within a resolved-flow volume cell, the constructive interference of all ray volumes can lead to a total GW signal with a negative vertical derivative in potential temperature larger than the positive derivative given by the stratification of the resolved flow. Once this threshold is exceeded turbulence is invoked, causing  turbulent viscosity and diffusivity that provide a dissipative right-hand side to the spectral wave-action equation (\ref{equation_wact_cons_sp_lag_dim}) so that the GW field is kept at the threshold of static instability.

The interaction between the parameterized GWs and the resolved large-scale flow takes two directions. The contribution of the large-scale winds $\langle\uv\rangle$ to the group velocity $\cg$ and to the wavenumber velocity $\kvp$ influences the development of the wave-action density. In the other direction, GWs influence the resolved flow in its thermodynamics via the divergence of the GW entropy flux (\ref{eq_entrflux_dim}). They also change its momentum via the divergence of GW momentum flux (\ref{eq_momflux_uu_dim}) - (\ref{eq_momflux_wv_dim}), and via the elastic term, i.e. the last term in (\ref{eq_hormmomeq_dim}). That term can be obtained from the GW entropy-flux convergence as well. The wavenumber integrals in the fluxes are presently estimated to first-order accuracy, by evaluating the integrand at the center of the ray volume and multiplying it by its wavenumber volume. So far implementations have been into finite-volume dynamical model cores for resolved-flow dynamics. In this context the fluxes are then projected onto the finite-volume cell faces and finally used there. 

It would seem attractive to avoid the projection of the fluxes from the Lagrangian GW model onto the resolved-flow finite volume cells, and also the interpolation of the resolved-flow winds to the location of each ray volume, by a straightforward finite-volume implementation of the spectral wave-action equation (\ref{equation_wact_cons_sp_dim}) in flux form. This alternative has been studied by \cite{Boeloenietal2016}. It turns out that the Lagrangian approach is computationally much more efficient. Two factors contribute to this: Firstly, in a finite-volume approach, one must span a six-dimensional phase space, with often a substantial fraction of cells not contributing essentially to the GW fluxes. Secondly, GW refraction and reflection are only captured well in the finite-volume approach if the resolution in wavenumber space is excessively high.

\subsection{\label{ss_gwp_class}Comparison with Classic GWP}

The approach described above on the simulation of the interaction between subgrid-scale GWs and a resolved flow, in climate models but also in weather-forecast codes, differs in various regards from the approach presently applied by the weather and climate centers: First, we point out that present-day GW parameterizations use the observation that the GW impact on the resolved-flow QGPV in (\ref{eq_pv_igwimp_dim}) could also be obtained by dropping the GW entropy-flux convergence from the mean-flow entropy equation (\ref{eq_synpteq_dim}), by removing the elastic term from the mean-flow momentum equation (\ref{eq_hormmomeq_dim}), and by replacing the momentum flux there by the pseudo-momentum flux, i.e. by using 
\begin{equation}
  \label{eq_synpteq_pmappr_dim}
  \left(\partial_t + \langle\vv\rangle \cdot \nabla\right) \langle\theta\rangle = 0
\end{equation}
and
\begin{eqnarray}
  && 
  \left(\partial_t + \langle\vv\rangle \cdot \nabla\right) \vua + f \ez \times \vua \nonumber\\
  &&= 
  - c_p \langle\theta\rangle \nabla_h \pia - \frac{1}{\rhobar} \nabla \cdot \left(\vGc\ex + \vHc\ey\right)
  \label{eq_hormmomeq_pmappr_dim}
\end{eqnarray}
and then starting from there the asymptotic analysis of synoptic-scale flow dynamics. Therefore the common approach is to only force the resolved-flow momentum equation, by GW pseudomomentum-flux convergence instead of GW momentum-flux convergence, and to not have the GW parameterization acting on the thermodynamics. It is computationally simpler and also has a certain conceptual attractiveness because all of the GW impact can be attributed to the pseudomomentum-flux divergence. From the theory, this could be a viable approach, but it assumes that the resolved flow is geostrophically and hydrostatically balanced. Given the general trend to increasingly finer model grids, with also larger-scale unbalanced GWs more and more being part of the resolved flow, this is however an issue that has been investigated by \cite{Weietal2019}. Indeed they show, in comparisons between idealized GW resolving simulations and coarse-grid simulations with a GW parameterization either using the general (direct) approach or the classic (pseudomomentum) approach, that the former is much more reliable in the simulation of the GW-mean-flow interaction.

The second and third aspect relate to simplifications in the implementation that are mostly due to considerations of computational efficiency: Because code parallelization typically works in vertical columns, i.e. different processors are allotted different horizontal locations, it is computationally less expensive to use parameterizations for subgrid-scale processes that do not couple different horizontal positions. In the context here this amounts to ignoring the impact of horizontal mean-flow gradients on the horizontal wavenumbers, i.e. assuming $\dot{k} = \dot{l} = 0$, and to ignoring all horizontal group-velocity components, and hence to replacing the spectral wave-action equations (\ref{equation_wact_cons_sp_dim}) and (\ref{equation_wact_cons_sp_lag_dim}) by
\begin{eqnarray}
  \label{equation_wact_cons_sp_sc}
  \partial_t\Nc + \partial_z \left(c_{gz} \Nc\right) + \partial_m \left(\dot{m} \Nc\right) &=& 0\\
  \label{equation_wact_cons_sp_lag_sc}
  \partial_t \Nc + c_{gz} \partial_z \Nc + \dot{m} \partial_m \Nc &=& 0
\end{eqnarray}
This neglects horizontal GW propagation, and it also neglects the response of the horizontal wave number to horizontal variations of the resolved flow. Moreover, in this so-called single-column approximation one also neglects horizontal GW fluxes in the GW impact on the large-scale flow, i.e. one approximates (\ref{eq_hormmomeq_pmappr_dim}) by 
\begin{eqnarray}
  && 
  \left(\partial_t + \langle\vv\rangle \cdot \nabla\right) \vua + f \ez \times \vua \nonumber\\
  &&= 
  - c_p \langle\theta\rangle \nabla_h \pia - \frac{1}{\rhobar} \partial_z \left(\Gc_z\ex + \Hc_z\ey\right)
  \label{eq_hormmomeq_pmappr_sc}
\end{eqnarray}

Finally, in a further attempt to gain in efficiency, one neglects, in the so-called steady-state approximation, the time dependence of the wave-action density, i.e. one uses instead of (\ref{equation_wact_cons_sp_sc})
\begin{equation}
  \label{equation_wact_cons_sp_ss}
  \partial_z \left(c_{gz} \Nc\right) + \partial_m \left(\dot{m} \Nc\right) = 0
\end{equation}
or, integrating in vertical wavenumber
\begin{equation}
  \label{equation_wact_cons_sp_ss_a}
  \partial_z \int dm\, c_{gz} \Nc = 0
\end{equation}
Given a lower boundary condition, obtained from some description of the GW sources the latter equation is integrated vertically to obtain a profile of wave-action density that agrees with what one would get, from a steady lower boundary condition, after a period sufficiently long for rays to propagate from the lower boundary to the model top. Effectively this is then an equilibrium profile in agreement with instantaneous GW propagation from the lower boundary to the model top. In practice, one always has a discrete sum of spectral components that are emitted from some parameterized source. Hence one recurs to the representation (\ref{equation_wact_sp_dim}) and solves the single-column and steady-state simplification 
\begin{equation}
    \partial_z \left(c_{gz,\beta} \Ac_\beta\right) = 0
\end{equation}
of (\ref{eq_waveact}) for each spectral component.

As follows from the non-acceleration result discussed in subsection \ref{ss_na} the classic approach using the single-column and steady-state approximations, the first neglecting GW horizontal propagation and horizontal GW fluxes, and the second neglecting GW transience, relies exclusively on GW dissipation as a process allowing a GW impact on the resolved flow. The relevance of GW transience has been investigated by \cite{Boeloenietal2016,Boeloenietal2021,Kimetal2021}. As shown by \cite{Boeloenietal2016}, the resolved-flow impact induced by the breaking of a single GW packet is often described quite incorrectly if GW transience is not taken into account. The consequence of this for GW parameterization in a global climate model has been studied by \cite{Boeloenietal2021,Kimetal2021}. The result is that the statistics of simulated GW fluxes, exhibiting in measurements distributions with long tails of strong fluxes, are affected significantly by a steady-state assumption, to the point that the distributions are distorted significantly away from the observational findings. 

The consequences of the single-column approximation have been investigated in the mean time as well (Völker et al 2023, Kim et al 2023, both to be submitted elsewhere). As with regard to the neglect of wave transience, they appear to be of leading order as well. The horizontal distribution of GW fluxes in the middle atmosphere (above 15km altitude) is very different between the single-column and the general parameterization approach so that also the large-scale winds differ significantly. In the tropics this leads to a significantly changed variability, with the quasi-biennial oscillation \cite[e.g.]{baldwinetal2021} differing in its period conspicuously between simulations using the two approaches. Hence it appears that the general approach is considerably more realistic.

\section{\label{sec_discussion} Final discussion}

The multi-scale theory outlined above has led to the development of an extended approach, in weather and climate models, for the parameterization of subgrid-scale GWs. As is clear by now it is considerably more realistic than classic GW parameterizations, where the GW impact has been simplified by focusing on GW pseudomomentum, where GW transience is neglected, and where the effects of horizontal GW propagation and of horizontal GW fluxes are neglected. This comes at a computational prize: Simulations with the new approach are slower, by about an order of magnitude at global-model resolutions tested so far, than those using a classic GW parameterization. Yet, they are more realistic, and if one wanted to resolve the GWs that turn out to matter, simulations would be and are \cite[e.g.]{Stephanetal2020,Stephanetal2022} by many orders of magnitude slower than even this \cite{Boeloenietal2021}. Hence the approach discussed here seems to be a reasonable compromise between the realism of GW permitting simulations, certainly of their own value as a data source for many studies, and the efficiency of climate simulations using classic GW parameterizations. Moreover, in view of the increasing complexity of ever higher resolved weather and climate models \cite{Slingoetal2022} there is an ever-increasing need for a hierarchy of models that help us gain conceptual insight into the complex atmosphere \cite{held2005}. In this hierarchy the more general approach for GW parameterizations should have its place.

Open issues remain, both on the theoretical and on the applied side. One aspect that seems to deserve consideration is the interaction between GWs and the geostrophic vortical mode that is the other constituting component of mesoscale atmospheric dynamics \cite[e.g.]{Calliesetal2014,Achatzetal2017}. Present approaches for the description of this interaction, using tools of wave-turbulence theory \cite{Edenetal2019}, do not take the presence of a leading-order mean flow into account. In the ocean context this is possible, but in the atmosphere mean winds enter to leading order. This also holds for another aspect of relevance, the so-far neglect of GW-GW interactions. It is not clear how relevant this process will be in the end, and to the best of our knowledge this has not been investigated yet. Here as well available theories \cite{Hasselmann1966} suffer from the neglect of a mean flow. It would be interesting to obtain a corresponding extension of the theory, e.g. following the route indicated by \cite{Voelkeretal2021}. Such work would close a conceptual gap that we still seem to have in the weakly nonlinear regime between the quasilinear regime (allowing a spectral approach) and the large-amplitude regime (where only locally monochromatic GW fields are possible). Finally, the handling of GW breaking by the saturation approach is very crude and a theory encompassing GW-turbulence interaction still needs to be derived from the basic equations. Similar considerations hold for the emission of GWs by various processes, be it flow over orography, convection, or emission by jets and fronts, to just name the most often discussed candidates. In all of these instances closed mathematical descriptions are still to be derived from multi-scale approaches, that hopefully would be able to replace present schemes. Finally, we think that in the numerical implementation of the general approach the last word has not been spoken. It would be helpful if experts in numerical mathematics gave it a closer look, especially with regard to accuracy and efficiency.

\added[id=UA]{Finally we also want to point out that it might be of interest to consider and extend the tools and techniques outlined here for other challenging problems of wave-mean-flow-interaction theory. The quasi-biennial oscillation of the equatorial zonal-mean zonal winds \cite{baldwinetal2021}, e.g., is not only due to GWs but also to larger-scale tropical Kelvin and Rossby-Gravity waves. Moreover, Kelvin waves also interact with GWs in the tropics \cite{Kim_Achatz_2021_Interaction}. The interaction between mesoscale GWs, larger-scale tropical waves and the zonal-mean flow could be an interesting problem to be studied using these methods. Another field of application could be caustics in general, e.g. in nonlinear acoustics \cite{Hunter_Keller_1984_Wave}, quantum mechanics \citep{Gosse2006}, general relativity \citep{Manor1977}, or plasma physics \citep{Pereverzev1992}, where the spectral approach outlined here could help the development of closed and comparatively simple treatments.}

\begin{acknowledgments}
UA thanks the German Research Foundation (DFG) for partial support through the research unit "Multiscale Dynamics of Gravity Waves" (MS-GWaves, grants Grants AC 71/8-2, AC 71/9-2, and AC 71/12-2) and CRC 301 "TPChange" (Project-ID 428312742, Projects B06 “Impact of small-scale dynamics on UTLS transport and mixing” and B07 “Impact of cirrus clouds on tropopause structure”). 
YHK and UA thank the German Federal Ministry of Education and Research (BMBF) for partial support through the program Role of the Middle Atmosphere in Climate (ROMIC II: QUBICC) and through grant 01LG1905B.
UA and GSV thank the  German Research Foundation (DFG) for partial support through the CRC 181 “Energy transfers in Atmosphere an Ocean” (Project Number 274762653, Projects W01  “Gravity-wave parameterization for the atmosphere” and S02 “Improved Parameterizations and Numerics in Climate Models.”).
UA is furthermore grateful for support by Eric and Wendy Schmidt through the Schmidt Futures VESRI “DataWave” project.
\end{acknowledgments}


\providecommand{\noopsort}[1]{}\providecommand{\singleletter}[1]{#1}%

\end{document}